\documentclass[aps,prb,twocolumn,superscriptaddress,showpacs]{revtex4-1}
\usepackage{graphicx,bm,mathtools}

\begin{document}

\title{Evidence for a hybridization gap in noncentrosymmetric CeRuSi$_3$}
\date{\today}

\author{M. Smidman}
\affiliation{Department of Physics, University of Warwick, Coventry CV4 7AL, United Kingdom}
\author{D. T. Adroja}
\email[]{devashibhai.adroja@stfc.ac.uk}
\affiliation{ISIS Facility, STFC, Rutherford Appleton Laboratory, Chilton, Didcot, Oxfordshire OX11 0QX, United Kingdom}
\affiliation{Highly Correlated Matter Research Group, Physics Department, University of Johannesburg, Auckland Park 2006, South Africa}
\author{E. A. Goremychkin}
\affiliation{ISIS Facility, STFC, Rutherford Appleton Laboratory, Chilton, Didcot, Oxfordshire OX11 0QX, United Kingdom}
\author{M. R. Lees}
\affiliation{Department of Physics, University of Warwick, Coventry CV4 7AL, United Kingdom}
\author{D. M. Paul}
\affiliation{Department of Physics, University of Warwick, Coventry CV4 7AL, United Kingdom}
\author{G. Balakrishnan}
\email[]{g.balakrishnan@warwick.ac.uk}
\affiliation{Department of Physics, University of Warwick, Coventry CV4 7AL, United Kingdom}

\date{\today}

\begin{abstract}
Inelastic neutron scattering (INS) and specific heat measurements have been performed on the intermediate valence compound CeRuSi$_3$, which is isostructural to the noncentrosymmetric pressure-induced superconductors  CeRhSi$_3$,  CeIrSi$_3$ and  CeCoGe$_3$. INS measurements at 7~K reveal a broad peak at (58.5~$\pm$~1.4)~meV, while at 300~K, broad quasielastic scattering is observed. This indicates a large Kondo temperature of $T_{\rm K}~\sim~680$~K. The magnetic contribution to the specific heat ($C_{\rm mag}$) has a value of $\gamma~=~62.5(1)$~mJ/mol~K$^2$ at low temperatures and above about 100~K can be well accounted for by the Coqlbin-Schrieffer model with a characteristic temperature of $T_0$~=~680~K, which is further evidence that CeRuSi$_3$ is in the intermediate valence regime.

\end{abstract}


\maketitle

\section{Introduction}

There has been considerable recent interest in the coexistence of superconductivity and magnetism in heavy fermion compounds. In particular, there are several heavy fermion systems where the electronic ground state can be tuned via a non-thermal parameter such as doping, pressure or magnetic fields \cite{NatureMagSC}. These systems are often explained in the framework of the Doniach phase diagram \cite{Doniach1977}, due to competition between the intersite Ruderman-Kittel-Kasuya-Yosida (RKKY) interaction, which leads to magnetic ordering of the cerium $4f$ electrons and the onsite Kondo interaction, which favors a non-magnetic ground state. In the region where magnetic order is suppressed, a superconducting dome is often observed \cite{PfleidererRMP}. Magnetic fluctuations are believed to play an important role in mediating the superconductivity in both heavy fermion and high temperature superconductors.

Several members of the Ce$TX_3$ ($T$~=~transition metal, $X$~=~Si or Ge) series display antiferromagnetic order but become superconducting at sufficiently large pressures. These compounds crystallize in the noncentrosymmetric, tetragonal BaNiSn$_3$ type structure (space group $I4mm$). Due to lack of inversion symmetry and a finite antisymmetric spin-orbit coupling, the spin degeneracy of the conduction bands is lifted and the superconducting state is a mixture of spin singlet and triplet states \cite{NCSGorkov,BauerNCS}. For example, at ambient pressure CeRhSi$_3$ orders antiferromagnetically at 1.6~K but becomes superconducting for $p~>$~1.2~GPa \cite{CeRhSi3Rep,CeRhSi3SC}. The superconducting state displays large and anisotropic upper critical fields, reaching up to 30~T for magnetic fields along the $c$~axis \cite{CeRhSi3Hc2}. 

\begin{figure}[tb]
\begin{center}
  \includegraphics[width=0.75\columnwidth]{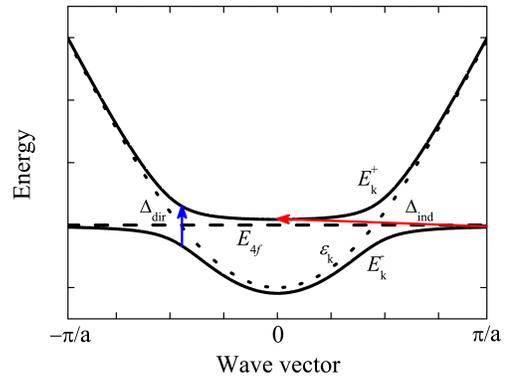}
\end{center}
	\caption{A schematic diagram of the hybridized bands $E^+_k$ and  $E^-_k$ of the Kondo lattice which result from the hybridization of the conduction and 4$f$ electrons, following Ref.~\onlinecite{HFSemi}. The unhybridized conduction ($\epsilon_k$) and 4$f$  ($E_{4f}$) bands are shown by the dashed and dotted lines respectively. A finite hybridization opens a gap between the bands and the direct ($\Delta_{\rm dir}$) and indirect ($\Delta_{\rm ind}$) gaps are indicated by arrows.}
   \label{HFBandss}
\end{figure}

It is therefore desirable to characterize Ce$TX_3$ compounds at ambient pressure, to examine the role played by the competition between the RKKY and Kondo interactions in determining the ground state properties and to identify the proximity of the system to quantum criticality. However, several compounds such as CeFeGe$_3$ \cite{CeFeGe3Rep}, CeRuSi$_3$ \cite{CeTX32008} and CeCoSi$_3$ are reported to have non-magnetic ground states and a broad maximum is observed in the magnetic susceptibility at around 50, 150 and 200~K respectively. In these systems, the hybridization between the $4f$ and conduction electrons is believed to be strong enough that the Kondo interaction is dominant and magnetic order is entirely suppressed. Within the Kondo lattice model, strong hybridization leads to the formation of renormalized bands and a hybridization gap, shown schematically in Fig.~\ref{HFBandss} \cite{SpinGapRev,HFSemi}. If the Fermi level lies within the gap, insulating behavior in the resistivity is observed, while if it lies on top of the lower hybridized band, the system is metallic despite the presence of a gap. For sufficiently strong hybridization, the system would be expected to enter the intermediate valence regime and this has been observed at low temperatures in resonant inverse photoemission spectroscopy measurements of the CeCoGe$_{1-x}$Si$_x$ system \cite{CeCoGe31999}.

Although inelastic neutron scattering (INS) investigations have been performed on several Ce$T$X$_3$ compounds for ($T$~=~Co, Rh and Ir, $X$~=~Si, Ge) \cite{CeCoGe3MS, CeRhGe32012, AdrojaCeTSi3}, as far as we are aware there are no reports of such measurements on isostructural CeRuSi$_3$. The aim of the present work is to investigate the magnetic INS response of CeRuSi$_3$ to probe the nature of the $4f$ electrons, that is whether they display localized or itinerant behavior. The existence of a hybridization gap at low temperatures is deduced and this is studied using INS and specific heat measurements. Furthermore, this study will provide an ideal comparison with existing INS data of Ce$T$Si$_3$ compounds and characterizing the strength of the hybridization between the 4$f$ and conduction electrons may allow a greater understanding of the role of this hybridization in determining the nature of the ground state.

\section{Experimental Details}

Polycrystalline samples of CeRuSi$_3$ and LaRuSi$_3$ were prepared by arc-melting stoichiometric quantities of the constituent elements on a water cooled copper hearth, in an argon atmosphere. The samples were flipped and melted several times to improve homogeneity and then wrapped in Ta foil, sealed in evacuated quartz tubes and annealed at 900~$^{\circ}$C for two weeks. Powder x-ray diffraction measurements were performed using a  Bruker D5005 diffractometer. Lattice parameters of $a~=~4.2106(2)$ and $4.2597(3)$~\AA, and $c~=~9.9204(7)$ and $9.9382(9)$~\AA~were obtained for CeRuSi$_3$ and LaRuSi$_3$ respectively. A small number of impurity peaks are present in the patterns of both compounds, the largest of which has an intensity of $\sim$4\% of the largest sample peak. Inelastic neutron scattering measurements were performed on the MERLIN time-of-flight spectrometer in the ISIS facility at the Rutherford Appleton Laboratory, U.K. The samples were wrapped in Al foil and cooled to 7~K in a closed cycle refrigerator in He-exchange gas, to thermalize the sample temperature. Measurements were made at 7 and 300~K with $E_{\rm i}$~= 30 and 200~meV selected via a Fermi chopper. Specific heat measurements were performed between 1.8 and 350~K using the two-tau relaxation method with a Quantum Design Physical Property Measurement System (PPMS).

\section{Results and Discussion}
\subsection{Inelastic neutron scattering}

\begin{figure}[tb]
\begin{center}
  \includegraphics[width=0.99\columnwidth]{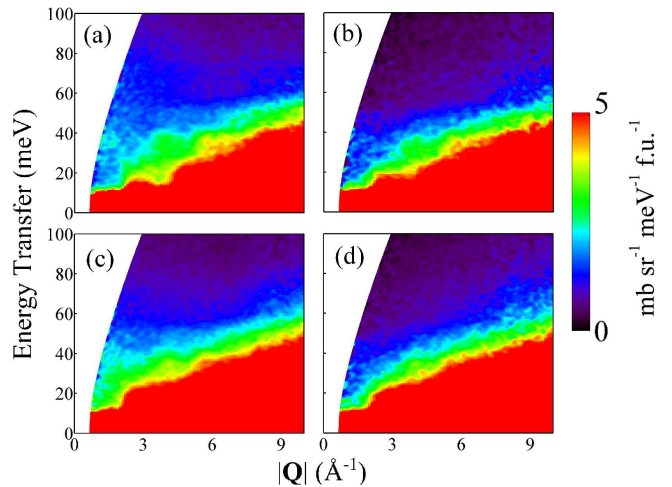}
\end{center}
	\caption{Colour-coded plots of the INS intensity with $E_{\rm{i}}$~=~200~meV for (a) CeRuSi$_3$ at 7~K, (b) LaRuSi$_3$ at 7~K, (c) CeRuSi$_3$ at 300~K. (d) LaRuSi$_3$ at 300~K.}
   \label{CeRuSi3Colour}
\end{figure}

Colour-coded plots of the INS intensity [$S(Q, \omega)$] in absolute units of mb~sr$^{-1}$~meV$^{-1}$~f.u.$^{-1}$ (f.u.~=~formula units) for $E_{\rm i}$~=~200~meV are shown in Fig.~\ref{CeRuSi3Colour}. The scattering of CeRuSi$_3$ and LaRuSi$_3$ at 7~K are shown in Figs.~\ref{CeRuSi3Colour}(a) and (b) respectively. Extra intensity at low $\mathbf{|Q|}$ can be identified in the CeRuSi$_3$ plot, which is evidence for a magnetic contribution to the scattering. Although this is most intense at around 50~meV, it extends up close to 100~meV, whereas the scattering is negligible for LaRuSi$_3$ at these energy transfers, apart from at high  $\mathbf{|Q|}$. The scattering of LaRuSi$_3$ at 300~K (Fig.~\ref{CeRuSi3Colour}(d)) is similar to that observed at low temperatures. However, although magnetic scattering can still be identified in the measurements of CeRuSi$_3$ at 300~K (Fig.~\ref{CeRuSi3Colour}(c)), it has shifted to lower energies. The low $\mathbf{|Q|}$ scattering is significantly reduced above 50~meV, suggesting a change in behaviour with increasing temperature.

\begin{figure}[tb]
\begin{center}
\includegraphics[width=0.99\columnwidth]{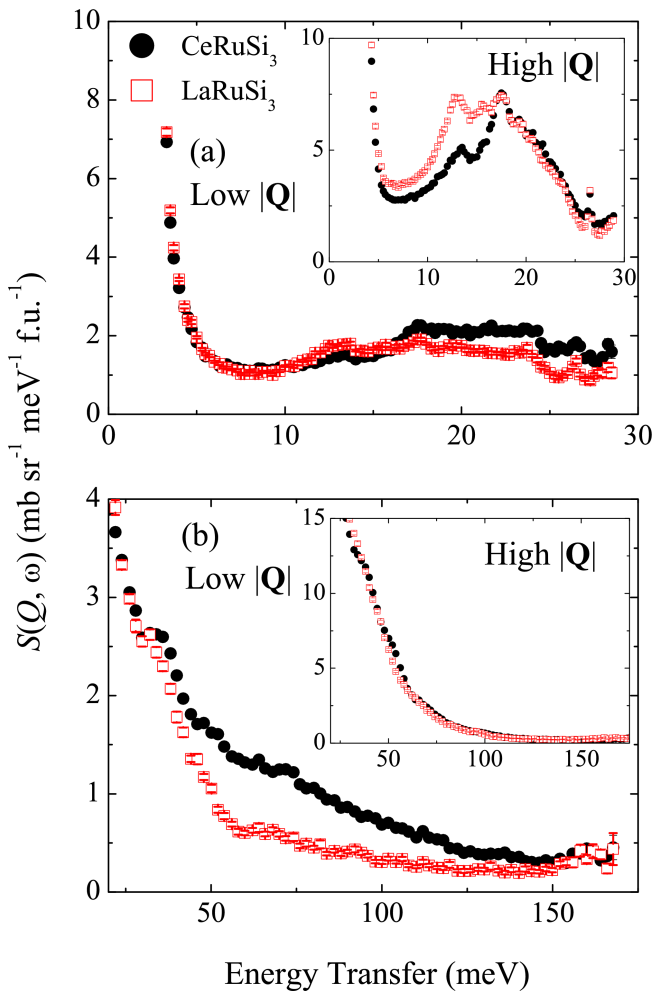}
\end{center}
	\caption{Low $\mathbf{|Q|}$ cuts of $S(Q, \omega)$ for CeRuSi$_3$ and LaRuSi$_3$ at 7~K with incident energies of (a) 30~meV and (b) 200~meV, with the corresponding high $\mathbf{|Q|}$ cuts shown in the insets. The low $\mathbf{|Q|}$ cuts were integrated from 0-4~\AA$^{-1}$\space and 0-6~\AA$^{-1}$, while the high $\mathbf{|Q|}$ cuts were integrated from 4-7~\AA$^{-1}$\space and 12-18~\AA$^{-1}$\space for  $E_{\rm i}$~=~30 and 200~meV respectively.}
\label{CeRuSi3Cuts}
\end{figure}

Cuts of $S(Q, \omega)$ were made by integrating across low and high values of $\mathbf{|Q|}$. Low $\mathbf{|Q|}$ cuts are displayed for $E_{\rm i}$~=~30 and 200~meV in Figs.~\ref{CeRuSi3Cuts}(a) and (b) respectively and the corresponding high $\mathbf{|Q|}$ cuts are shown in the insets. In Fig.~\ref{CeRuSi3Cuts}(a), it can be seen that for energy transfers greater than about 17~meV, the low  $\mathbf{|Q|}$ scattering of CeRuSi$_3$ is greater than that of  LaRuSi$_3$, whereas the high  $\mathbf{|Q|}$ scattering is similar for both compounds. Similarly, the low $\mathbf{|Q|}$ scattering of CeRuSi$_3$ in  Fig.~\ref{CeRuSi3Cuts}(b) is greater than that of LaRuSi$_3$ for energy transfers greater than 35~meV, up to at least 150~meV. This is strong evidence for the presence of a magnetic contribution to the low $\mathbf{|Q|}$ scattering of CeRuSi$_3$, which is not observed at high  $\mathbf{|Q|}$. This is because the magnetic intensity is proportional to the square of the magnetic form factor ($F(\mathbf{Q})$), which is maximum at $\mathbf{|Q|}~=~0$ and has a negligible contribution for $\mathbf{|Q|}~>~10$~\AA$^{-1}$.  The high $\mathbf{|Q|}$ scattering is similar in both compounds, which  is expected since at large momentum transfers, $S(Q, \omega)$ is dominated by phonon scattering, which increases as $Q^2$. An exception to this is at energy transfers between 5 and 15~meV, displayed in the inset of Fig.~\ref{CeRuSi3Cuts}(a). Although both plots of the high $\mathbf{|Q|}$ scattering display a peak at around 13~meV, the LaRuSi$_3$ data is of a greater intensity at these energy transfers. This is likely because the coherent scattering cross section of La is approximately three times that of Ce and this indicates that the phonon scattering in this region is dominated by the rare earth atoms.

\begin{figure}[tb]
\begin{center}
   \includegraphics[width=0.8\columnwidth]{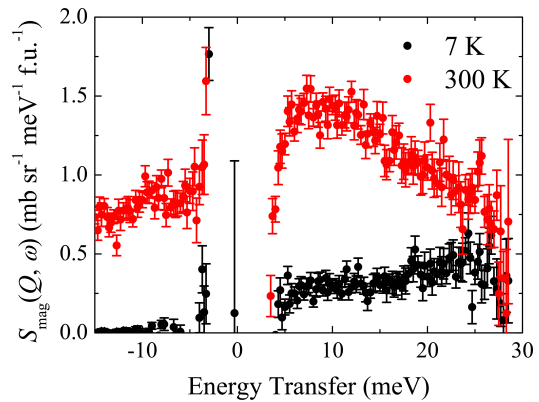}
\end{center}
	\caption{Magnetic scattering of CeRuSi$_3$ at 7 and 300~K for $E_{\rm i}$=~30~meV,~ estimated using Eq.~\ref{SmagEq}.}
\label{CeRuSi3Mag30meV}
\end{figure}

The magnetic contribution to the scattering [$S_{\rm mag}(Q, \omega)$] was obtained by subtracting an estimate of the phonon scattering. This is commonly estimated by directly subtracting the spectra of the non-magnetic reference compound from that of the magnetic compound. However, this method does not adequately remove the contributions from phonon scattering. Instead, the ratio between the high $(S^{\rm hQ}_{\rm ph}$) and low ($S^{\rm lQ}_{\rm ph}$) $\mathbf{|Q|}$ cuts was calculated for the reference compound LaRuSi$_3$. If the phonon scattering of the magnetic scattering scales similarly with $\mathbf{|Q|}$, the magnetic scattering can be estimated from the low $\mathbf{|Q|}$  ($S^{\rm lQ}_{\rm CeRuSi_3}$) and high $\mathbf{|Q|}$  ($S^{\rm hQ}_{\rm CeRuSi_3}$) scattering using

\begin{equation}
S_{\rm mag} = S^{\rm lQ}_{\rm CeRuSi_3} - \frac{S^{\rm lQ}_{\rm ph}}{S^{\rm hQ}_{\rm ph}}S^{\rm hQ}_{\rm CeRuSi_3}.
\label{SmagEq}
\end{equation}

\begin{figure}[tb]
\begin{center}
   \includegraphics[width=0.8\columnwidth]{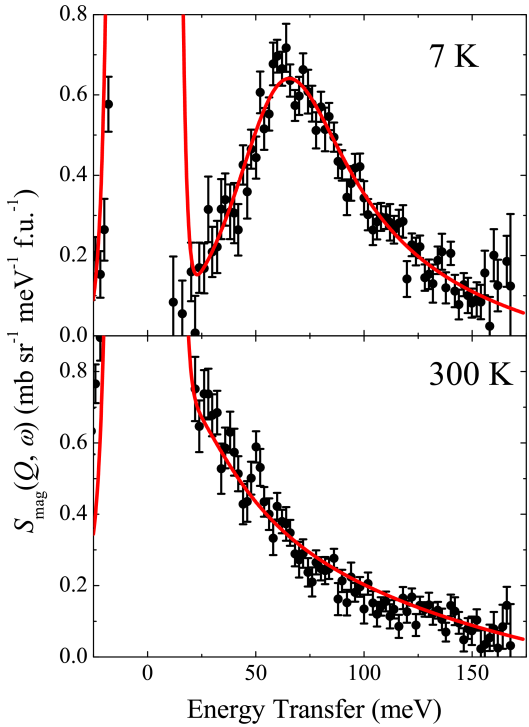}
\end{center}
	\caption{Magnetic scattering of CeRuSi$_3$ at 7 and 300~K for $E_{\rm i}$~=~200~meV, estimated using Eq.~\ref{SmagEq}.}
\label{CeRuSi3Smag200meV}
\end{figure}

\noindent $S_{\rm mag}(Q, \omega)$ is displayed in Fig.~\ref{CeRuSi3Mag30meV} for $E_{\rm i}$~=~30~meV. At 300~K a broad quasielastic response is observed, whereas there is no quasielastic scattering at 7~K and the magnetic scattering is greatly reduced. At 7~K, the magnetic scattering gradually increases with increasing energy transfers up to at least 25~meV and this may be the tail of an excitation centred at higher energies. The magnetic scattering  for $E_{\rm i}$~=~200~meV is displayed in Fig.~\ref{CeRuSi3Smag200meV}, where  broad quasielastic scattering is also observed at 300~K. However at 7~K, the magnetic response has shifted to a broad peak centred at energy transfers of around 60~meV. The scattering at 7~K was fitted with a Lorentzian function convoluted with the instrument resolution. A peak center of (58.5~$\pm$~1.4)~meV with a linewidth (half-width at half maximum) of (31.9~$\pm$~1.3)~meV was obtained.  Using the second sum rule $\int S_{\rm mag}(\omega)/F^2(Q)~{\rm d}\omega~=~48.8\mu_{\rm eff}^2$, an effective moment of $\mu_{\rm eff}~=~1.49~\mu_{\rm B}$ is calculated. Although this is lower than the expected value of $2.54~\mu_{\rm B}$, this result indicates the bulk origin of the magnetic scattering and this suggests there may be an additional magnetic response at energy transfers greater than 170~meV. For the 300~K data, the quasielastic scattering was fitted with a Lorentzian function, centred on the position of the elastic line. A linewidth of 30(3)~meV is obtained, which is similar to the width of the inelastic excitation at 7~K. The observed INS response of an inelastic peak at low temperatures and quasielastic scattering at high temperatures is typical behavior for intermediate valence compounds \cite{SpinGapRev}. The lack of both quasielastic scattering and well defined crystalline electric field (CEF) excitations is also expected for these materials. The position of the inelastic peak  indicates the scale of the  Kondo temperature \cite{CePd3INSPoly}, giving $T_{\rm K}~\sim~680$~K. This is smaller than that of the isostructural, intermediate valence compound CeCoSi$_3$, where $T_{\rm K}~\sim~990(50)$~K is estimated from a peak in the magnetic INS scattering at 85~meV  \cite{AdrojaCeTSi3}, while $T_{\rm K}$ of the order of 900~K is estimated from specific heat and magnetic susceptibility measurements  \cite{CeCoGe31998}. However, $T_{\rm K}$ is larger than for the magnetically ordered CeRhSi$_3$ and CeIrSi$_3$, which has been estimated to be around 100~K for both compounds \cite{CeRhSi3Rep}. It should also be noted that the threshold for the onset of magnetic scattering is predicted to be 0.8 times the peak energy \cite{CeFe4Sb122007}, but no clear threshold is observed. The increase of the magnetic scattering at around 20~meV in Fig.~\ref{CeRuSi3Mag30meV} suggests that scattering from the low energy tail of the peak near 60~meV is present down to at least these energy transfers.

\subsection{Specific heat}

\begin{figure}[tb]
\begin{center}
   \includegraphics[width=0.8\columnwidth]{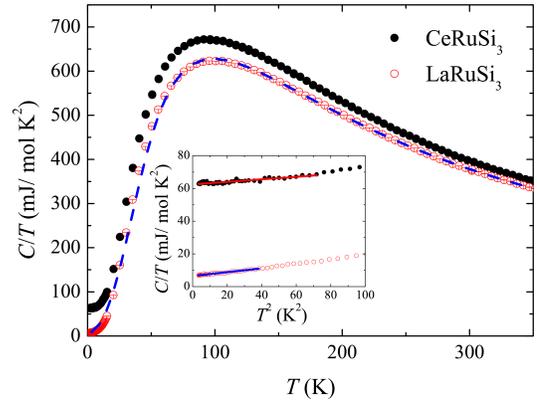}
\end{center}
	\caption{Temperature dependence of the specific heat ($C$) of CeRuSi$_3$ and LaRuSi$_3$. The dashed line shows a fit to the specific heat of LaRuSi$_3$ described in the text. The inset shows plots of $C/T$ against $T^2$ with linear fits for both compounds. }
\label{CeRuSi3C}
\end{figure}

The specific heats of CeRuSi$_3$ and LaRuSi$_3$ are displayed in Fig.~\ref{CeRuSi3C}. The specific heat of CeRuSi$_3$ is consistently larger than that of LaRuSi$_3$, indicating the presence of a magnetic contribution. As shown in the inset, the residual value of $C/T$ at zero temperature is larger in CeRuSi$_3$. A linear fit to $C/T$ against $T^2$ gives $\gamma~=~62.5(1)$~mJ/mol~K$^2$ for CeRuSi$_3$, while a smaller value of ~6.5(1)~mJ/mol~K$^2$ is obtained for LaRuSi$_3$, indicating an enhancement of $\gamma$ due to hybridization between 4$f$ and conduction electrons.

\begin{figure}[tb]
\begin{center}
   \includegraphics[trim=1cm 0.4cm 1.3cm 2.3cm, clip=true, width=0.95\columnwidth]{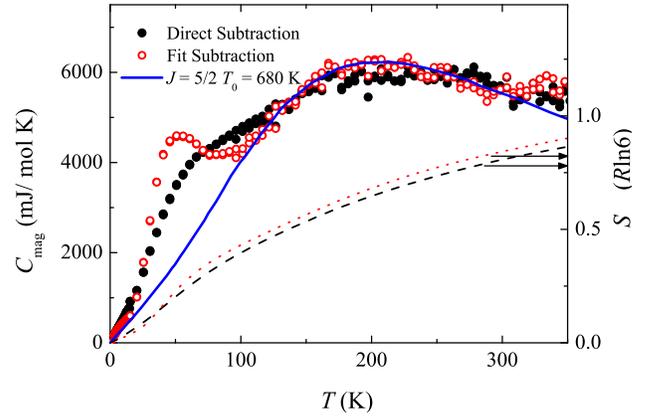}
\end{center}
	\caption{Magnetic contribution to the specific heat ($C_{\rm mag}$)  of CeRuSi$_3$ estimated by a direct subtraction of the specific heat of LaRuSi$_3$ and of a fit to the LaRuSi$_3$ data described in the text. The solid line shows the specific heat of Coqlbin-Schrieffer model for $J~=~\frac{5}{2}$ from  Ref.~\onlinecite{RajanModel}, with a characteristic temperature of $T_0~=~680$~K. The temperature dependence of the entropy in units of $R$ln6 are shown by the dashed and dotted lines for the direct and fit subtractions respectively. }
\label{CeRuSi3Cmag}
\end{figure}

The magnetic contribution to the specific heat from the 4$f$ electrons ($C_{\rm mag}$) was calculated by subtracting the specific heat of LaRuSi$_3$ by two methods. Firstly, the specific heat of LaRuSi$_3$ was directly subtracted from that of CeRuSi$_3$. Secondly, the specific heat of LaRuSi$_3$ was fitted with the sum of $\gamma T$ and an estimate of the phonon contribution ($C_{\rm ph}$), where  $\gamma$ was fixed to the low temperature value. $C_{\rm ph}$ was fitted using the sum of a Debye and Einstein term \cite{Kittel}. A Debye temperature of $\theta_D~=~314$~K and an Einstein temperature of  $\theta_E~=~569$~K were obtained, with a  $73\%$ contribution from the Debye term. This is displayed by the dashed line in Fig.~\ref{CeRuSi3C} and shows that the specific heat of LaRuSi$_3$ can be accounted for with the sum of a phonon contribution and a linear electronic term.  The two estimates of $C_{\rm mag}$ are displayed in Fig.~\ref{CeRuSi3Cmag}. In both plots, $C_{\rm mag}$ displays a broad peak at around 200~K.  An additional feature is observed at around 50-60~K. In the estimate from the fit subtraction, there is a peak in this region while in the direct subtraction, there is a clear shoulder. The dotted and dashed lines show the temperature dependence of the entropy ($S_{\rm mag}$) in units of $R$ln(6), obtained from integrating $C_{\rm mag}/T$ from the fit and direct subtractions respectively. The entropy reaches $0.9R$ln(6) and $0.87R$ln(6) at 350~K for the respective plots. 

The solid line shows the Coqlbin-Schrieffer model for $J~=~\frac{5}{2}$ as calculated by Rajan in Ref.~\onlinecite{RajanModel}. The one adjustable parameter in the model is the characteristic temperature $T_0$, which was fixed to 680~K, which corresponds to the energy of the INS peak position. The model accounts well for the data at high temperatures. However below temperatures of about 100~K, there is poor agreement with the data and there is an additional contribution at low temperatures, including the feature at around 50-60~K. This suggests there may be an additional low energy scale in the specific heat. It might be expected that this could lead to a two gaps  being observed in INS measurements, as is the case for CeFe$_2$Al$_{10}$ \cite{CeFe2Al10INS}.  If two gaps are present in CeRuSi$_3$, the weak rise in magnetic scattering above 5~meV at 7~K for $E_{\rm i}$=~30~meV  shown in Fig.~\ref{CeRuSi3Mag30meV} may be due to transitions across the lower energy gap.  It would also be interesting to investigate the gap structure of CeRuSi$_3$ with optical conductivity measurements, which would provide information on the size of the direct or charge gap. The ratio of the charge to the spin gap can provide important information on the dimensionality of the magnetic interactions \cite{SpinGapRev}.

\section{Conclusions}

INS and specific heat measurements have been performed on CeRuSi$_3$ and LaRuSi$_3$ and the results are consistent with CeRuSi$_3$ being an intermediate valence compound. The INS measurements of CeRuSi$_3$ at 7~K indicate the lack of both quasielastic scattering and narrow excitations associated with transitions between CEF levels. Instead a broad inelastic peak is observed, centered at (58.5~$\pm$~1.4)~meV with a linewidth of (31.9~$\pm$~1.3)~meV. The opening of such a hybridization gap in the inelastic response which transforms to quasielastic scattering at higher temperatures, indicates that CeRuSi$_3$ is an intermediate valence compound with $T_{\rm K}~\sim~680$~K. 

The magnetic contribution to the specific heat shows a broad peak at around  200~K and above temperatures of around 100~K, there is good agreement  with the  Coqlbin-Schrieffer model, with $T_0$~=~680~K. However at lower temperatures there is an additional contribution to the specific heat, which may indicate the presence of a lower energy scale in the compound. The value of  $\gamma$ is 62.5(1)~mJ/mol~K$^2$  for CeRuSi$_3$ compared to 6.5(1)~mJ/mol~K$^2$ for LaRuSi$_3$. This is a moderate enhancement due to electronic correlations and is comparable to the values observed for the isostructural CeRhGe$_3$ \cite{CeRhGe32012} and CeCoGe$_3$ \cite{CeCoGe32005}. Both CeRhGe$_3$ and CeCoGe$_3$  order antiferromagnetically and INS measurements reveal well defined CEF levels. The different ground state behavior of CeRuSi$_3$ results from the larger value of $T_{\rm K}$, which is an order of magnitude smaller, at around 10~K in the aforementioned ordered compounds \cite{CeRhGe32012,CeCoGe3MS}. The value of $\gamma$ is also larger than that observed in CeCoSi$_3$ of 37~mJ/mol~K$^2$  and is similar
to CeCoGe$_{0.75}$Si$_{2.25}$ \cite{CeCoGe31998}, which is in closer proximity to quantum criticality. 

There is particular interest in determining whether the properties of intermediate valence compounds can be accounted for by the single-ion, Anderson impurity model or whether coherence effects arising from the regular arrangement of rare earth ions needs to be taken into account. In the case of the latter, the system is described by the Anderson lattice model and at low temperatures, renormalized bands with a hybridization gap and large effective masses are expected to emerge. A $\mathbf{Q}$ dependence of the magnetic scattering would be strong evidence for coherent behavior, as observed in CePd$_3$ \cite{CePd3INSCrys}. It would therefore be interesting to study the INS response of single crystals of   CeRuSi$_3$  and to compare the results to both the Anderson impurity and Anderson lattice models.

\begin{acknowledgments}
We acknowledge the EPSRC, UK for providing funding (grant number EP/I007210/1). DTA thanks CMPC-STFC (grant number CMPC-09108) for financial support We thank T.E. Orton for technical support. Some of the equipment used in this research at the University of Warwick was obtained through the Science City Advanced Materials: Creating and Characterising Next Generation Advanced Materials Project, with support from Advantage West Midlands (AWM) and part funded by the European Regional Development Fund (ERDF). 
\end{acknowledgments}

\bibliography{CeRuSi3a}

\begin{thebibliography}{24}%
\makeatletter
\providecommand \@ifxundefined [1]{%
 \@ifx{#1\undefined}
}%
\providecommand \@ifnum [1]{%
 \ifnum #1\expandafter \@firstoftwo
 \else \expandafter \@secondoftwo
 \fi
}%
\providecommand \@ifx [1]{%
 \ifx #1\expandafter \@firstoftwo
 \else \expandafter \@secondoftwo
 \fi
}%
\providecommand \natexlab [1]{#1}%
\providecommand \enquote  [1]{``#1''}%
\providecommand \bibnamefont  [1]{#1}%
\providecommand \bibfnamefont [1]{#1}%
\providecommand \citenamefont [1]{#1}%
\providecommand \href@noop [0]{\@secondoftwo}%
\providecommand \href [0]{\begingroup \@sanitize@url \@href}%
\providecommand \@href[1]{\@@startlink{#1}\@@href}%
\providecommand \@@href[1]{\endgroup#1\@@endlink}%
\providecommand \@sanitize@url [0]{\catcode `\\12\catcode `\$12\catcode
  `\&12\catcode `\#12\catcode `\^12\catcode `\_12\catcode `\%12\relax}%
\providecommand \@@startlink[1]{}%
\providecommand \@@endlink[0]{}%
\providecommand \url  [0]{\begingroup\@sanitize@url \@url }%
\providecommand \@url [1]{\endgroup\@href {#1}{\urlprefix }}%
\providecommand \urlprefix  [0]{URL }%
\providecommand \Eprint [0]{\href }%
\providecommand \doibase [0]{http://dx.doi.org/}%
\providecommand \selectlanguage [0]{\@gobble}%
\providecommand \bibinfo  [0]{\@secondoftwo}%
\providecommand \bibfield  [0]{\@secondoftwo}%
\providecommand \translation [1]{[#1]}%
\providecommand \BibitemOpen [0]{}%
\providecommand \bibitemStop [0]{}%
\providecommand \bibitemNoStop [0]{.\EOS\space}%
\providecommand \EOS [0]{\spacefactor3000\relax}%
\providecommand \BibitemShut  [1]{\csname bibitem#1\endcsname}%
\let\auto@bib@innerbib\@empty
\bibitem [{\citenamefont {Mathur}\ \emph {et~al.}(1998)\citenamefont {Mathur},
  \citenamefont {Grosche}, \citenamefont {Julian}, \citenamefont {Walker},
  \citenamefont {Freye}, \citenamefont {Haswelwimmer},\ and\ \citenamefont
  {Lonzarich}}]{NatureMagSC}%
  \BibitemOpen
  \bibfield  {author} {\bibinfo {author} {\bibfnamefont {N.~D.}\ \bibnamefont
  {Mathur}}, \bibinfo {author} {\bibfnamefont {F.~M.}\ \bibnamefont {Grosche}},
  \bibinfo {author} {\bibfnamefont {S.~R.}\ \bibnamefont {Julian}}, \bibinfo
  {author} {\bibfnamefont {I.~R.}\ \bibnamefont {Walker}}, \bibinfo {author}
  {\bibfnamefont {D.}~\bibnamefont {Freye}}, \bibinfo {author} {\bibfnamefont
  {R.~K.~W.}\ \bibnamefont {Haswelwimmer}}, \ and\ \bibinfo {author}
  {\bibfnamefont {G.~G.}\ \bibnamefont {Lonzarich}},\ }\href@noop {} {\bibfield
   {journal} {\bibinfo  {journal} {Nature}\ }\textbf {\bibinfo {volume}
  {394}},\ \bibinfo {pages} {39} (\bibinfo {year} {1998})}\BibitemShut
  {NoStop}%
\bibitem [{\citenamefont {Doniach}(1977)}]{Doniach1977}%
  \BibitemOpen
  \bibfield  {author} {\bibinfo {author} {\bibfnamefont {S.}~\bibnamefont
  {Doniach}},\ }\href {\doibase 10.1016/0378-4363(77)90190-5} {\bibfield
  {journal} {\bibinfo  {journal} {Physica B+C}\ }\textbf {\bibinfo {volume}
  {91}},\ \bibinfo {pages} {231 } (\bibinfo {year} {1977})}\BibitemShut
  {NoStop}%
\bibitem [{\citenamefont {Pfleiderer}(2009)}]{PfleidererRMP}%
  \BibitemOpen
  \bibfield  {author} {\bibinfo {author} {\bibfnamefont {C.}~\bibnamefont
  {Pfleiderer}},\ }\href {\doibase 10.1103/RevModPhys.81.1551} {\bibfield
  {journal} {\bibinfo  {journal} {Rev. Mod. Phys.}\ }\textbf {\bibinfo {volume}
  {81}},\ \bibinfo {pages} {1551} (\bibinfo {year} {2009})}\BibitemShut
  {NoStop}%
\bibitem [{\citenamefont {Gor'kov}\ and\ \citenamefont
  {Rashba}(2001)}]{NCSGorkov}%
  \BibitemOpen
  \bibfield  {author} {\bibinfo {author} {\bibfnamefont {L.~P.}\ \bibnamefont
  {Gor'kov}}\ and\ \bibinfo {author} {\bibfnamefont {E.~I.}\ \bibnamefont
  {Rashba}},\ }\href {\doibase 10.1103/PhysRevLett.87.037004} {\bibfield
  {journal} {\bibinfo  {journal} {Phys. Rev. Lett.}\ }\textbf {\bibinfo
  {volume} {87}},\ \bibinfo {pages} {037004} (\bibinfo {year}
  {2001})}\BibitemShut {NoStop}%
\bibitem [{\citenamefont {Bauer}\ and\ \citenamefont
  {Sigrist}(2012)}]{BauerNCS}%
  \BibitemOpen
  \bibfield  {author} {\bibinfo {author} {\bibfnamefont {E.}~\bibnamefont
  {Bauer}}\ and\ \bibinfo {author} {\bibfnamefont {M.}~\bibnamefont
  {Sigrist}},\ }\href {http://books.google.co.uk/books?id=nDZ4lKD00t8C} {\emph
  {\bibinfo {title} {Non-Centrosymmetric Superconductors: Introduction and
  Overview}}},\ Lecture notes in physics\ (\bibinfo  {publisher}
  {Springer-Verlag Berlin Heidelberg},\ \bibinfo {year} {2012})\BibitemShut
  {NoStop}%
\bibitem [{\citenamefont {Muro}\ \emph {et~al.}(1998)\citenamefont {Muro},
  \citenamefont {Eom}, \citenamefont {Takeda},\ and\ \citenamefont
  {Ishikawa}}]{CeRhSi3Rep}%
  \BibitemOpen
  \bibfield  {author} {\bibinfo {author} {\bibfnamefont {Y.}~\bibnamefont
  {Muro}}, \bibinfo {author} {\bibfnamefont {D.}~\bibnamefont {Eom}}, \bibinfo
  {author} {\bibfnamefont {N.}~\bibnamefont {Takeda}}, \ and\ \bibinfo {author}
  {\bibfnamefont {M.}~\bibnamefont {Ishikawa}},\ }\href {\doibase
  10.1143/JPSJ.67.3601} {\bibfield  {journal} {\bibinfo  {journal} {J. Phys.
  Soc. Jpn.}\ }\textbf {\bibinfo {volume} {67}},\ \bibinfo {pages} {3601}
  (\bibinfo {year} {1998})}\BibitemShut {NoStop}%
\bibitem [{\citenamefont {Kimura}\ \emph {et~al.}(2005)\citenamefont {Kimura},
  \citenamefont {Ito}, \citenamefont {Saitoh}, \citenamefont {Umeda},
  \citenamefont {Aoki},\ and\ \citenamefont {Terashima}}]{CeRhSi3SC}%
  \BibitemOpen
  \bibfield  {author} {\bibinfo {author} {\bibfnamefont {N.}~\bibnamefont
  {Kimura}}, \bibinfo {author} {\bibfnamefont {K.}~\bibnamefont {Ito}},
  \bibinfo {author} {\bibfnamefont {K.}~\bibnamefont {Saitoh}}, \bibinfo
  {author} {\bibfnamefont {Y.}~\bibnamefont {Umeda}}, \bibinfo {author}
  {\bibfnamefont {H.}~\bibnamefont {Aoki}}, \ and\ \bibinfo {author}
  {\bibfnamefont {T.}~\bibnamefont {Terashima}},\ }\href {\doibase
  10.1103/PhysRevLett.95.247004} {\bibfield  {journal} {\bibinfo  {journal}
  {Phys. Rev. Lett.}\ }\textbf {\bibinfo {volume} {95}},\ \bibinfo {pages}
  {247004} (\bibinfo {year} {2005})}\BibitemShut {NoStop}%
\bibitem [{\citenamefont {Kimura}\ \emph {et~al.}(2007)\citenamefont {Kimura},
  \citenamefont {Ito}, \citenamefont {Aoki}, \citenamefont {Uji},\ and\
  \citenamefont {Terashima}}]{CeRhSi3Hc2}%
  \BibitemOpen
  \bibfield  {author} {\bibinfo {author} {\bibfnamefont {N.}~\bibnamefont
  {Kimura}}, \bibinfo {author} {\bibfnamefont {K.}~\bibnamefont {Ito}},
  \bibinfo {author} {\bibfnamefont {H.}~\bibnamefont {Aoki}}, \bibinfo {author}
  {\bibfnamefont {S.}~\bibnamefont {Uji}}, \ and\ \bibinfo {author}
  {\bibfnamefont {T.}~\bibnamefont {Terashima}},\ }\href {\doibase
  10.1103/PhysRevLett.98.197001} {\bibfield  {journal} {\bibinfo  {journal}
  {Phys. Rev. Lett.}\ }\textbf {\bibinfo {volume} {98}},\ \bibinfo {pages}
  {197001} (\bibinfo {year} {2007})}\BibitemShut {NoStop}%
\bibitem [{\citenamefont {Riseborough}(2000)}]{HFSemi}%
  \BibitemOpen
  \bibfield  {author} {\bibinfo {author} {\bibfnamefont {P.~S.}\ \bibnamefont
  {Riseborough}},\ }\href {\doibase 10.1080/000187300243345} {\bibfield
  {journal} {\bibinfo  {journal} {Adv. Phys.}\ }\textbf {\bibinfo {volume}
  {49}},\ \bibinfo {pages} {257} (\bibinfo {year} {2000})}\BibitemShut
  {NoStop}%
\bibitem [{\citenamefont {Yamamoto}\ \emph {et~al.}(1995)\citenamefont
  {Yamamoto}, \citenamefont {Ishikawa}, \citenamefont {Hasegawa},\ and\
  \citenamefont {Sakurai}}]{CeFeGe3Rep}%
  \BibitemOpen
  \bibfield  {author} {\bibinfo {author} {\bibfnamefont {H.}~\bibnamefont
  {Yamamoto}}, \bibinfo {author} {\bibfnamefont {M.}~\bibnamefont {Ishikawa}},
  \bibinfo {author} {\bibfnamefont {K.}~\bibnamefont {Hasegawa}}, \ and\
  \bibinfo {author} {\bibfnamefont {J.}~\bibnamefont {Sakurai}},\ }\href
  {\doibase 10.1103/PhysRevB.52.10136} {\bibfield  {journal} {\bibinfo
  {journal} {Phys. Rev. B}\ }\textbf {\bibinfo {volume} {52}},\ \bibinfo
  {pages} {10136} (\bibinfo {year} {1995})}\BibitemShut {NoStop}%
\bibitem [{\citenamefont {Kawai}\ \emph {et~al.}(2008)\citenamefont {Kawai},
  \citenamefont {Muranaka}, \citenamefont {Measson}, \citenamefont {Shimoda},
  \citenamefont {Doi}, \citenamefont {Matsuda}, \citenamefont {Haga},
  \citenamefont {Knebel}, \citenamefont {Lapertot}, \citenamefont {Aoki},
  \citenamefont {Flouquet}, \citenamefont {Takeuchi}, \citenamefont {Settai},\
  and\ \citenamefont {\={O}nuki}}]{CeTX32008}%
  \BibitemOpen
  \bibfield  {author} {\bibinfo {author} {\bibfnamefont {T.}~\bibnamefont
  {Kawai}}, \bibinfo {author} {\bibfnamefont {H.}~\bibnamefont {Muranaka}},
  \bibinfo {author} {\bibfnamefont {M.-A.}\ \bibnamefont {Measson}}, \bibinfo
  {author} {\bibfnamefont {T.}~\bibnamefont {Shimoda}}, \bibinfo {author}
  {\bibfnamefont {Y.}~\bibnamefont {Doi}}, \bibinfo {author} {\bibfnamefont
  {T.~D.}\ \bibnamefont {Matsuda}}, \bibinfo {author} {\bibfnamefont
  {Y.}~\bibnamefont {Haga}}, \bibinfo {author} {\bibfnamefont {G.}~\bibnamefont
  {Knebel}}, \bibinfo {author} {\bibfnamefont {G.}~\bibnamefont {Lapertot}},
  \bibinfo {author} {\bibfnamefont {D.}~\bibnamefont {Aoki}}, \bibinfo {author}
  {\bibfnamefont {J.}~\bibnamefont {Flouquet}}, \bibinfo {author}
  {\bibfnamefont {T.}~\bibnamefont {Takeuchi}}, \bibinfo {author}
  {\bibfnamefont {R.}~\bibnamefont {Settai}}, \ and\ \bibinfo {author}
  {\bibfnamefont {Y.}~\bibnamefont {\={O}nuki}},\ }\href {\doibase
  10.1143/JPSJ.77.064716} {\bibfield  {journal} {\bibinfo  {journal} {J. Phys.
  Soc. Jpn.}\ }\textbf {\bibinfo {volume} {77}},\ \bibinfo {pages} {064716}
  (\bibinfo {year} {2008})}\BibitemShut {NoStop}%
\bibitem [{\citenamefont {Adroja}\ \emph {et~al.}(2008)\citenamefont {Adroja},
  \citenamefont {McEwen}, \citenamefont {Park}, \citenamefont {Hillier},
  \citenamefont {Takeda}, \citenamefont {Riseborough},\ and\ \citenamefont
  {Takabatake}}]{SpinGapRev}%
  \BibitemOpen
  \bibfield  {author} {\bibinfo {author} {\bibfnamefont {D.~T.}\ \bibnamefont
  {Adroja}}, \bibinfo {author} {\bibfnamefont {K.~A.}\ \bibnamefont {McEwen}},
  \bibinfo {author} {\bibfnamefont {J.~G.}\ \bibnamefont {Park}}, \bibinfo
  {author} {\bibfnamefont {A.~D.}\ \bibnamefont {Hillier}}, \bibinfo {author}
  {\bibfnamefont {N.}~\bibnamefont {Takeda}}, \bibinfo {author} {\bibfnamefont
  {P.~S.}\ \bibnamefont {Riseborough}}, \ and\ \bibinfo {author} {\bibfnamefont
  {T.}~\bibnamefont {Takabatake}},\ }\href@noop {} {\bibfield  {journal}
  {\bibinfo  {journal} {J. Optoelectron. Adv. M.}\ }\textbf {\bibinfo {volume}
  {10}},\ \bibinfo {pages} {1564 } (\bibinfo {year} {2008})}\BibitemShut
  {NoStop}%
\bibitem [{\citenamefont {Kanai}\ \emph {et~al.}(1999)\citenamefont {Kanai},
  \citenamefont {Terashima}, \citenamefont {Eom}, \citenamefont {Ishikawa},\
  and\ \citenamefont {Shin}}]{CeCoGe31999}%
  \BibitemOpen
  \bibfield  {author} {\bibinfo {author} {\bibfnamefont {K.}~\bibnamefont
  {Kanai}}, \bibinfo {author} {\bibfnamefont {T.}~\bibnamefont {Terashima}},
  \bibinfo {author} {\bibfnamefont {D.~H.}\ \bibnamefont {Eom}}, \bibinfo
  {author} {\bibfnamefont {M.}~\bibnamefont {Ishikawa}}, \ and\ \bibinfo
  {author} {\bibfnamefont {S.}~\bibnamefont {Shin}},\ }\href {\doibase
  10.1103/PhysRevB.60.R9900} {\bibfield  {journal} {\bibinfo  {journal} {Phys.
  Rev. B}\ }\textbf {\bibinfo {volume} {60}},\ \bibinfo {pages} {R9900}
  (\bibinfo {year} {1999})}\BibitemShut {NoStop}%
\bibitem [{\citenamefont {Smidman}\ \emph {et~al.}(2013)\citenamefont
  {Smidman}, \citenamefont {Adroja}, \citenamefont {Hillier}, \citenamefont
  {Chapon}, \citenamefont {Taylor}, \citenamefont {Anand}, \citenamefont
  {Singh}, \citenamefont {Lees}, \citenamefont {Goremychkin}, \citenamefont
  {Koza}, \citenamefont {Krishnamurthy}, \citenamefont {Paul},\ and\
  \citenamefont {Balakrishnan}}]{CeCoGe3MS}%
  \BibitemOpen
  \bibfield  {author} {\bibinfo {author} {\bibfnamefont {M.}~\bibnamefont
  {Smidman}}, \bibinfo {author} {\bibfnamefont {D.~T.}\ \bibnamefont {Adroja}},
  \bibinfo {author} {\bibfnamefont {A.~D.}\ \bibnamefont {Hillier}}, \bibinfo
  {author} {\bibfnamefont {L.~C.}\ \bibnamefont {Chapon}}, \bibinfo {author}
  {\bibfnamefont {J.~W.}\ \bibnamefont {Taylor}}, \bibinfo {author}
  {\bibfnamefont {V.~K.}\ \bibnamefont {Anand}}, \bibinfo {author}
  {\bibfnamefont {R.~P.}\ \bibnamefont {Singh}}, \bibinfo {author}
  {\bibfnamefont {M.~R.}\ \bibnamefont {Lees}}, \bibinfo {author}
  {\bibfnamefont {E.~A.}\ \bibnamefont {Goremychkin}}, \bibinfo {author}
  {\bibfnamefont {M.~M.}\ \bibnamefont {Koza}}, \bibinfo {author}
  {\bibfnamefont {V.~V.}\ \bibnamefont {Krishnamurthy}}, \bibinfo {author}
  {\bibfnamefont {D.~M.}\ \bibnamefont {Paul}}, \ and\ \bibinfo {author}
  {\bibfnamefont {G.}~\bibnamefont {Balakrishnan}},\ }\href {\doibase
  10.1103/PhysRevB.88.134416} {\bibfield  {journal} {\bibinfo  {journal} {Phys.
  Rev. B}\ }\textbf {\bibinfo {volume} {88}},\ \bibinfo {pages} {134416}
  (\bibinfo {year} {2013})}\BibitemShut {NoStop}%
\bibitem [{\citenamefont {Hillier}\ \emph {et~al.}(2012)\citenamefont
  {Hillier}, \citenamefont {Adroja}, \citenamefont {Manuel}, \citenamefont
  {Anand}, \citenamefont {Taylor}, \citenamefont {McEwen}, \citenamefont
  {Rainford},\ and\ \citenamefont {Koza}}]{CeRhGe32012}%
  \BibitemOpen
  \bibfield  {author} {\bibinfo {author} {\bibfnamefont {A.~D.}\ \bibnamefont
  {Hillier}}, \bibinfo {author} {\bibfnamefont {D.~T.}\ \bibnamefont {Adroja}},
  \bibinfo {author} {\bibfnamefont {P.}~\bibnamefont {Manuel}}, \bibinfo
  {author} {\bibfnamefont {V.~K.}\ \bibnamefont {Anand}}, \bibinfo {author}
  {\bibfnamefont {J.~W.}\ \bibnamefont {Taylor}}, \bibinfo {author}
  {\bibfnamefont {K.~A.}\ \bibnamefont {McEwen}}, \bibinfo {author}
  {\bibfnamefont {B.~D.}\ \bibnamefont {Rainford}}, \ and\ \bibinfo {author}
  {\bibfnamefont {M.~M.}\ \bibnamefont {Koza}},\ }\href {\doibase
  10.1103/PhysRevB.85.134405} {\bibfield  {journal} {\bibinfo  {journal} {Phys.
  Rev. B}\ }\textbf {\bibinfo {volume} {85}},\ \bibinfo {pages} {134405}
  (\bibinfo {year} {2012})}\BibitemShut {NoStop}%
\bibitem [{\citenamefont {Adroja}\ \emph {et~al.}()\citenamefont {Adroja} \emph
  {et~al.}}]{AdrojaCeTSi3}%
  \BibitemOpen
  \bibfield  {author} {\bibinfo {author} {\bibfnamefont {D.~T.}\ \bibnamefont
  {Adroja}} \emph {et~al.},\ }\href@noop {} {}\bibinfo {howpublished}
  {(unpublished)}\BibitemShut {NoStop}%
\bibitem [{\citenamefont {Murani}\ \emph {et~al.}(1996)\citenamefont {Murani},
  \citenamefont {Raphel}, \citenamefont {Bowden},\ and\ \citenamefont
  {Eccleston}}]{CePd3INSPoly}%
  \BibitemOpen
  \bibfield  {author} {\bibinfo {author} {\bibfnamefont {A.~P.}\ \bibnamefont
  {Murani}}, \bibinfo {author} {\bibfnamefont {R.}~\bibnamefont {Raphel}},
  \bibinfo {author} {\bibfnamefont {Z.~A.}\ \bibnamefont {Bowden}}, \ and\
  \bibinfo {author} {\bibfnamefont {R.~S.}\ \bibnamefont {Eccleston}},\ }\href
  {\doibase 10.1103/PhysRevB.53.8188} {\bibfield  {journal} {\bibinfo
  {journal} {Phys. Rev. B}\ }\textbf {\bibinfo {volume} {53}},\ \bibinfo
  {pages} {8188} (\bibinfo {year} {1996})}\BibitemShut {NoStop}%
\bibitem [{\citenamefont {Eom}\ \emph {et~al.}(1998)\citenamefont {Eom},
  \citenamefont {Ishikawa}, \citenamefont {Kitagawa},\ and\ \citenamefont
  {Takeda}}]{CeCoGe31998}%
  \BibitemOpen
  \bibfield  {author} {\bibinfo {author} {\bibfnamefont {D.}~\bibnamefont
  {Eom}}, \bibinfo {author} {\bibfnamefont {M.}~\bibnamefont {Ishikawa}},
  \bibinfo {author} {\bibfnamefont {J.}~\bibnamefont {Kitagawa}}, \ and\
  \bibinfo {author} {\bibfnamefont {N.}~\bibnamefont {Takeda}},\ }\href
  {\doibase 10.1143/JPSJ.67.2495} {\bibfield  {journal} {\bibinfo  {journal}
  {J. Phys. Soc. Jpn.}\ }\textbf {\bibinfo {volume} {67}},\ \bibinfo {pages}
  {2495} (\bibinfo {year} {1998})}\BibitemShut {NoStop}%
\bibitem [{\citenamefont {Viennois}\ \emph {et~al.}(2007)\citenamefont
  {Viennois}, \citenamefont {Girard}, \citenamefont {Chapon}, \citenamefont
  {Adroja}, \citenamefont {Bewley}, \citenamefont {Ravot}, \citenamefont
  {Riseborough},\ and\ \citenamefont {Paschen}}]{CeFe4Sb122007}%
  \BibitemOpen
  \bibfield  {author} {\bibinfo {author} {\bibfnamefont {R.}~\bibnamefont
  {Viennois}}, \bibinfo {author} {\bibfnamefont {L.}~\bibnamefont {Girard}},
  \bibinfo {author} {\bibfnamefont {L.~C.}\ \bibnamefont {Chapon}}, \bibinfo
  {author} {\bibfnamefont {D.~T.}\ \bibnamefont {Adroja}}, \bibinfo {author}
  {\bibfnamefont {R.~I.}\ \bibnamefont {Bewley}}, \bibinfo {author}
  {\bibfnamefont {D.}~\bibnamefont {Ravot}}, \bibinfo {author} {\bibfnamefont
  {P.~S.}\ \bibnamefont {Riseborough}}, \ and\ \bibinfo {author} {\bibfnamefont
  {S.}~\bibnamefont {Paschen}},\ }\href {\doibase 10.1103/PhysRevB.76.174438}
  {\bibfield  {journal} {\bibinfo  {journal} {Phys. Rev. B}\ }\textbf {\bibinfo
  {volume} {76}},\ \bibinfo {pages} {174438} (\bibinfo {year}
  {2007})}\BibitemShut {NoStop}%
\bibitem [{\citenamefont {Rajan}(1983)}]{RajanModel}%
  \BibitemOpen
  \bibfield  {author} {\bibinfo {author} {\bibfnamefont {V.}~\bibnamefont
  {Rajan}},\ }\href {\doibase 10.1103/PhysRevLett.51.308} {\bibfield  {journal}
  {\bibinfo  {journal} {Phys. Rev. Lett.}\ }\textbf {\bibinfo {volume} {51}},\
  \bibinfo {pages} {308} (\bibinfo {year} {1983})}\BibitemShut {NoStop}%
\bibitem [{\citenamefont {Kittel}(1996)}]{Kittel}%
  \BibitemOpen
  \bibfield  {author} {\bibinfo {author} {\bibfnamefont {C.}~\bibnamefont
  {Kittel}},\ }\href@noop {} {\emph {\bibinfo {title} {Introduction to Solid
  State Physics}}},\ \bibinfo {edition} {7th}\ ed.\ (\bibinfo  {publisher}
  {John Wiley and Sons, New York},\ \bibinfo {year} {1996})\BibitemShut
  {NoStop}%
\bibitem [{\citenamefont {Adroja}\ \emph {et~al.}(2013)\citenamefont {Adroja},
  \citenamefont {Hillier}, \citenamefont {Muro}, \citenamefont {Kajino},
  \citenamefont {Takabatake}, \citenamefont {Peratheepan}, \citenamefont
  {Strydom}, \citenamefont {Deen}, \citenamefont {Demmel}, \citenamefont
  {Stewart}, \citenamefont {Taylor}, \citenamefont {Smith}, \citenamefont
  {Ramos},\ and\ \citenamefont {Adams}}]{CeFe2Al10INS}%
  \BibitemOpen
  \bibfield  {author} {\bibinfo {author} {\bibfnamefont {D.~T.}\ \bibnamefont
  {Adroja}}, \bibinfo {author} {\bibfnamefont {A.~D.}\ \bibnamefont {Hillier}},
  \bibinfo {author} {\bibfnamefont {Y.}~\bibnamefont {Muro}}, \bibinfo {author}
  {\bibfnamefont {J.}~\bibnamefont {Kajino}}, \bibinfo {author} {\bibfnamefont
  {T.}~\bibnamefont {Takabatake}}, \bibinfo {author} {\bibfnamefont
  {P.}~\bibnamefont {Peratheepan}}, \bibinfo {author} {\bibfnamefont {A.~M.}\
  \bibnamefont {Strydom}}, \bibinfo {author} {\bibfnamefont {P.~P.}\
  \bibnamefont {Deen}}, \bibinfo {author} {\bibfnamefont {F.}~\bibnamefont
  {Demmel}}, \bibinfo {author} {\bibfnamefont {J.~R.}\ \bibnamefont {Stewart}},
  \bibinfo {author} {\bibfnamefont {J.~W.}\ \bibnamefont {Taylor}}, \bibinfo
  {author} {\bibfnamefont {R.~I.}\ \bibnamefont {Smith}}, \bibinfo {author}
  {\bibfnamefont {S.}~\bibnamefont {Ramos}}, \ and\ \bibinfo {author}
  {\bibfnamefont {M.~A.}\ \bibnamefont {Adams}},\ }\href {\doibase
  10.1103/PhysRevB.87.224415} {\bibfield  {journal} {\bibinfo  {journal} {Phys.
  Rev. B}\ }\textbf {\bibinfo {volume} {87}},\ \bibinfo {pages} {224415}
  (\bibinfo {year} {2013})}\BibitemShut {NoStop}%
\bibitem [{\citenamefont {Thamizhavel}\ \emph {et~al.}(2005)\citenamefont
  {Thamizhavel}, \citenamefont {Takeuchi}, \citenamefont {Matsuda},
  \citenamefont {Haga}, \citenamefont {Sugiyama}, \citenamefont {Settai},\ and\
  \citenamefont {\={O}nuki}}]{CeCoGe32005}%
  \BibitemOpen
  \bibfield  {author} {\bibinfo {author} {\bibfnamefont {A.}~\bibnamefont
  {Thamizhavel}}, \bibinfo {author} {\bibfnamefont {T.}~\bibnamefont
  {Takeuchi}}, \bibinfo {author} {\bibfnamefont {T.~D.}\ \bibnamefont
  {Matsuda}}, \bibinfo {author} {\bibfnamefont {Y.}~\bibnamefont {Haga}},
  \bibinfo {author} {\bibfnamefont {K.}~\bibnamefont {Sugiyama}}, \bibinfo
  {author} {\bibfnamefont {R.}~\bibnamefont {Settai}}, \ and\ \bibinfo {author}
  {\bibfnamefont {Y.}~\bibnamefont {\={O}nuki}},\ }\href {\doibase
  10.1143/JPSJ.74.1858} {\bibfield  {journal} {\bibinfo  {journal} {J. Phys.
  Soc. Jpn.}\ }\textbf {\bibinfo {volume} {74}},\ \bibinfo {pages} {1858}
  (\bibinfo {year} {2005})}\BibitemShut {NoStop}%
\bibitem [{\citenamefont {Fanelli}\ \emph {et~al.}(2014)\citenamefont
  {Fanelli}, \citenamefont {Lawrence}, \citenamefont {Goremychkin},
  \citenamefont {Osborn}, \citenamefont {Bauer}, \citenamefont {McClellan},
  \citenamefont {Thompson}, \citenamefont {Booth}, \citenamefont
  {Christianson},\ and\ \citenamefont {Riseborough}}]{CePd3INSCrys}%
  \BibitemOpen
  \bibfield  {author} {\bibinfo {author} {\bibfnamefont {V.~R.}\ \bibnamefont
  {Fanelli}}, \bibinfo {author} {\bibfnamefont {J.~M.}\ \bibnamefont
  {Lawrence}}, \bibinfo {author} {\bibfnamefont {E.~A.}\ \bibnamefont
  {Goremychkin}}, \bibinfo {author} {\bibfnamefont {R.}~\bibnamefont {Osborn}},
  \bibinfo {author} {\bibfnamefont {E.~D.}\ \bibnamefont {Bauer}}, \bibinfo
  {author} {\bibfnamefont {K.~J.}\ \bibnamefont {McClellan}}, \bibinfo {author}
  {\bibfnamefont {J.~D.}\ \bibnamefont {Thompson}}, \bibinfo {author}
  {\bibfnamefont {C.~H.}\ \bibnamefont {Booth}}, \bibinfo {author}
  {\bibfnamefont {A.~D.}\ \bibnamefont {Christianson}}, \ and\ \bibinfo
  {author} {\bibfnamefont {P.~S.}\ \bibnamefont {Riseborough}},\ }\href
  {http://stacks.iop.org/0953-8984/26/i=22/a=225602} {\bibfield  {journal}
  {\bibinfo  {journal} {J. Phys. Condens. Matter}\ }\textbf {\bibinfo {volume}
  {26}},\ \bibinfo {pages} {225602} (\bibinfo {year} {2014})}\BibitemShut
  {NoStop}%
\end{thebibliography}%

\end{document}